\documentclass[aps, pra, superscriptaddress, twocolumn, amsmath,
amssymb, floatfix, 10pt]{revtex4-1}

\usepackage{graphicx}
\usepackage{siunitx}
\usepackage{quantum}
\usepackage{xfrac}
\usepackage{bm}
\usepackage{tikz}
\usepackage[colorlinks=false]{hyperref}
\usepackage{color}

\graphicspath{{./figures/}}

\newcommand{\infint}[1][]{\int_{-\infty}^{\infty} \!\!\! #1\,}
\DeclareMathOperator{\sinc}{sinc}
\DeclareMathOperator{\erfc}{erfc}

\begin{document}

\title{A source of polarization-entangled photon pairs interfacing
  quantum memories with telecom photons}

\author{Christoph Clausen}
\altaffiliation[Present address: ]{Vienna Center for Quantum Science
	and Technology, TU Wien - Atominstitut, Stadionallee 2,
	1020 Vienna, Austria.}
\email{christoph.clausen@ati.ac.at} 
\author{F\'{e}lix Bussi\`{e}res}
\email{felix.bussieres@unige.ch}
\author{Alexey Tiranov}
\affiliation{Group of Applied Physics, University of Geneva, CH-1211
  Geneva 4, Switzerland}

\author{Harald Herrmann}
\author{Christine Silberhorn}
\author{Wolfgang Sohler}
\affiliation{Applied Physics / Integrated Optics Group, University of
Paderborn, 33095 Paderborn, Germany}

\author{Mikael Afzelius}
\author{Nicolas Gisin}
\affiliation{Group of Applied Physics, University of Geneva, CH-1211
  Geneva 4, Switzerland}

\date{\today}

\begin{abstract}
  We present a source of polarization-entangled photon pairs suitable
  for the implementation of long-distance quantum communication
  protocols using quantum memories. Photon pairs with wavelengths
  \SI{883}{nm} and \SI{1338}{nm} are produced by coherently pumping
  two periodically poled nonlinear waveguides embedded in the arms of
  a polarization interferometer. Subsequent spectral filtering reduces
  the bandwidth of the photons to 240~MHz. The bandwidth is
  well-matched to a quantum memory based on an Nd:YSO crystal, to
  which, in addition, the center frequency of the \SI{883}{nm} photons
  is actively stabilized. A theoretical model that includes the
  effect of the filtering is presented and accurately fits the
  measured correlation functions of the generated photons. The model
  can also be used as a way to properly assess the properties of the
  source.  The quality of the entanglement is revealed by a visibility
  of $V=\SI{96.1(9)}{\%}$ in a Bell-type experiment and through the
  violation of a Bell inequality.
\end{abstract}

\maketitle

\section{Introduction}
Spontaneous parametric down-conversion (SPDC) is a simple and
efficient technique for the generation of non-classical light and of
photonic entanglement.
Several important tasks of quantum communication require photonic entanglement, but also optical quantum memories to store
this entanglement~\cite{Bussieres2013a}. A prominent example is the
quantum repeater~\cite{Briegel1998a,Sangouard2011a}, which can extend
the transmission distance of entanglement beyond the hard limit
dictated by loss in optical fibre. In this context, the combination of
photon pair sources and multimode quantum memories was
proposed~\cite{Simon2007}. The essence of this proposal is that the
sources create pairs comprised of one telecom-wavelength photon that
is used to distribute entanglement between distant nodes, while the
other photon is stored in a nearby quantum memory. This
increases the probability of successfully heralding a stored photon
when the telecom photon is detected. Multimode storage with selective
recall then multiplies the entanglement distribution rate by the
number of stored modes, and is essential to reach practical rates over
distances of 500~km or more~\cite{Sangouard2011a}.

Creating photon pairs such that one photon exactly matches the
absorption profile of the quantum memory, while the other is within a
telecom wavelength window of standard optical fibre, is a challenging
task in itself. Sources of photon pairs based on emissive atomic
ensembles or single emitters~\cite{Sangouard2011a} typically generate
photons at wavelengths in the vicinity of 800~nm, where the loss in
standard optical fibre is on the order $\sim \SI{3}{dB/km}$, i.e.~at least 10 times larger than
in telecom fibres. 
Reaching telecom wavelengths with such sources therefore requires frequency
conversion techniques, which has been
demonstrated~\cite{Ikuta2011a,Zaske2012a,DeGreve2012a,Pelc2012a,Albrecht2014},
but imposes an important technical overhead. SPDC offers much more
flexibility, since the wavelengths of the pump can be easily chosen
(and tuned) to directly generate the desired wavelengths. However,
unfiltered SPDC photons have a bandwidth on the order of hundreds of
GHz or more. Hence, they still need to be spectrally filtered to the
memory absorption bandwidth, which typically ranges from a few MHz to
a few GHz at most~\cite{Bussieres2013a}.

Different approaches for the filtering of SPDC photons were
demonstrated. Direct filtering (using Fabry-Perot cavities) of
frequency-degenerate photon pairs created in a lithium niobate
waveguide was first demonstrated~\cite{Akiba2009a}, and used for
storage of an heralded photon on the $D_1$ line (795~nm) of cold
rubidium atoms. The high conversion efficiency of the waveguide was
here used to counterbalance the extreme filtering (down to 9~MHz),
which effectively rejects almost all of the generated SPDC
bandwidth. A similar source was also developed to demonstrate the
heralded single-photon absorption by a single calcium atom at
854~nm~\cite{Piro2011}. Another approach is based on pumping a bulk
crystal put inside a cavity, yielding a doubly resonant optical
parametric oscillator (OPO) operated far below threshold. The cavity
effectively enhances the length of the nonlinear medium, and is
well-suited to generate narrowband photons. This was first
demonstrated with frequency-degenerate photons resonant with the $D_2$
line of rubidium (780~nm)~\cite{Bao2008,Zhang2011a}, and later with
photons resonant with the $D_1$ line (795~nm)~\cite{Scholz2009}. It
was also demonstrated with photon pairs generated at 1436 and
606~nm~\cite{Fekete2013}, and used for storage in a praseodymium-doped
crystal~\cite{Rielander2014a}. One important technical difficulty in
using an OPO is to fulfill the doubly resonant condition and
simultaneously lock one photon's frequency on the quantum memory. Even
though such sources can in principle emit the photons in a single
longitudinal mode with the help of the clustering
effect~\cite{Pomarico2009,Pomarico2012}, current state-of-the-art
sources~\cite{Foertsch2013,Fekete2013,Luo2013a} do not yet achieve all
the requiirements, and in practice some additional filtering outside
of the cavity is still necessary to remove spurious longitudinal
modes.

All the aforementioned experiments
produced photons with linewidths $\Delta \nu$ ranging from 1 to
20~MHz, which is dictated by the absorption bandwidth of the
respective quantum memory they were developed for. The coherence time
$\tau_c \sim 1/\Delta \nu$ of the photons produced can therefore be as
long as a microsecond, which impacts on the rate at which those
photons can be distributed. It is therefore desirable for the quantum
memory to absorb over a large bandwidth to increase the photon
distribution rate. 

In this article, we present a CW-pumped source of
polarization-entangled photon pairs with 240~MHz linewidth using a
direct filtering approach. This source was designed for experiments
involving quantum memories based on the atomic frequency comb protocol
(AFC)~\cite{Afzelius2009} in a Nd:YSO crystal. Earlier versions of
this source produced energy-time entangled photons with a smaller
linewidth, and was used to demonstrate the quantum storage of photonic
entanglement in a crystal~\cite{Clausen2011}, heralded entanglement of
two crystals~\cite{Usmani2012} and the storage of heralded
polarization qubits~\cite{Clausen2012}. Recently, the source described
in this paper was used to demonstrate the teleportation from a
telecom-wavelength photon to a solid-state quantum
memory~\cite{Bussieres2014}. We note that a similar source, based on a pulsed pump, was used 
for the storage of broadband time-bin entangled photons in a Tm:LiNbO$_3$ waveguide~\cite{Saglamyurek2011a}.

The paper is organized as follows. We give the requirements for the
photon-pair source in Sec.~\ref{sec:requirements}. The concept behind
the implementation is given in Sec.~\ref{sec:concept} with the details
of the actual implementation following in
Sec.~\ref{sec:implementation}. In
Sec.~\ref{sec:spectral-characterization} the spectral properties and
the correlation functions of the filtered photons are presented and
compared to the predictions of a model that includes the effect of the
filtering. The efficiency and detection rate of the source is
presented in section~\ref{sec:characterization}.
Section~\ref{sec:entanglement} presents measurements showing the high
degree of polarization entanglement of the photon pairs, as well as
its nonlocal nature. The appendices contain all the details pertaining
to the characterization of the source.

\section{Requirements}
\label{sec:requirements}
The source was designed for experiments involving an atomic frequency comb (AFC) type of quantum
memory in a Nd:YSO crystal, so the \emph{signal} photon of a pair has
to be in resonance with the transition from the ${}^4I_{9/2}$ ground
state to the ${}^4F_{3/2}$ excited state of the Nd$^{3+}$ ion at
$\lambda_s = \SI{883}{nm}$. Quantum communication over long distances
in optical fibre requires the wavelength of the idler photon of a pair
to be inside one of the so-called telecom windows, which span the
region from \SIrange{1300}{1700}{nm}. The condition for the idler
wavelength can be conveniently satisfied using a pump wavelength of
$\lambda_p = \SI{532}{nm}$, for which high-quality solid-state lasers
are readily available. This places the idler wavelength at $\lambda_i
= (\lambda_p^{-1} - \lambda_s^{-1})^{-1} = \SI{1338}{nm}$.

The bandwidth of the generated photon pairs is dictated by the
bandwidth of the quantum memory. In earlier experiments this bandwidth
was \SI{120}{MHz}~\cite{Clausen2011,Usmani2012}. Recently it has been
increased to about \SI{600}{MHz}~\cite{Bussieres2014}. Although this
is fairly large for a quantum memory, it is still 3 orders of
magnitude narrower than the typical bandwidth of photons generated by
SPDC, which is given by the phasematching condition and can be as
large as \SI{1}{THz}.

We also require quantum entanglement between the signal and idler
photons. Entanglement can be established between various degrees of
freedom. In particular energy-time entanglement is intrinsically
present when using a highly coherent pump laser. In this work, however,
we focus on polarization entanglement because of the experimental
convenience in manipulating and measuring the polarization state of
light.

\section{Concept}
\label{sec:concept}
Various schemes have been devised to generate polarization-entangled
photon pairs through SPDC\@. These schemes include selective
collection of photon pairs emitted at specific angles for
non-collinear \mbox{type-II} phasematching~\cite{Kwiat1995}, collinear
SPDC in two orthogonally oriented
crystals~\cite{Kwiat1999,Trojek2008}, and SPDC in Sagnac
interferometers~\cite{Kim2006, Hentschel2009}. We wanted to extend our
existing and well-functioning waveguide source~\cite{Clausen2011},
which is inherently collinear, to a configuration that can create
polarization-entangled photon pairs. Putting two waveguides back to
back is in principle possible, but as the cross-section of the
waveguides is only a few micrometres and may vary from
waveguide to waveguide, efficient and stable coupling from one to the
other is experimentally extremely challenging. Using a waveguide in a
Sagnac configuration is complicated by the need for achromatic optics
for coupling into and out of the waveguide and for the necessary
polarization rotation.

\begin{figure}
  \centering \def\svgwidth{\columnwidth}
  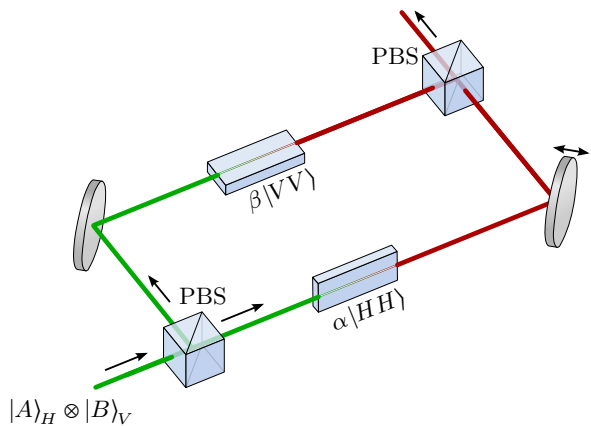
  \caption{Creation of
    polarization-entangled photon pairs with the help of two
    waveguides inside a polarization interferometer. A PBS coherently
    splits the pump photons according to their polarization. Each
    polarization component has a certain probability to be converted
    into a photon pair with the same polarization. The two
    polarization components of the photon pair are then recombined
    into the same spatial mode by a second PBS\@. The relative phase
    can be adjusted by moving one of the mirrors.}
  \label{fig:MachZehnder}
\end{figure}

To be able to efficiently employ our waveguides we follow the ideas
of~\cite{Kwiat1994,Kim2001} that suggest using a nonlinear crystal in
each arm of a polarization interferometer, as sketched in
Fig.~\ref{fig:MachZehnder}. We consider the situation of \mbox{type-I}
phasematching and that the two nonlinear crystals may have different
down-conversion efficiencies. Let the photons from the pump laser be
in a polarization state $\ket[H]{A}\otimes\ket[V]{B}$, where $\ket[H]{A}$
corresponds to a horizontally polarization coherent state of complex
amplitude $A$, and similarly for $\ket[V]{B}$. A polarizing beam
splitter (PBS) at the entrance of the interferometer splits the two
coherent state components in two paths. In the horizontal path the
photons can be converted into a photon pair $\ket{HH}$ with a
probability amplitude $\alpha \propto A$ by a first nonlinear
waveguide. A second waveguide rotated by \SI{90}{\degree} in the
vertical path can produce a photon pair $\ket{VV}$ with probability
amplitude $\beta \propto B$. Another PBS recombines the two paths, and
the final single-pair state $\ket{\psi_1}$ is given by
\begin{equation}
  \label{eq:photonpair}
  \ket{\psi_1} \propto \abs{\alpha}\ket{HH} +
  e^{i\phi}\abs{\beta}\ket{VV},
\end{equation}
where the phase $\phi$ depends on the path-length difference of the
interferometer, and on the relative phase between $\alpha$ and
$\beta$. By choosing the pump polarization such that it compensates
the efficiency difference, i.e.~$\abs{\alpha} = \abs{\beta}$,
and by slightly varying the position of one of the mirrors to obtain
$e^{i\phi} = \pm 1$, the single-pair state becomes equivalent to one
of the two Bell-states $\ket{\Phi_\pm} = (\ket{HH} \pm
\ket{VV})/\sqrt{2}$. However, one could equally well produce
non-maximally entangled states by choosing the polarization of the
pump laser accordingly.

\section{Implementation}
\label{sec:implementation}
In the following we detail the actual implementation of the source
of polarization-entangled photon pairs. We start by describing the two
waveguides that have been used. We then discuss the problem of
matching the spatial modes of the photons with the same wavelength
from different waveguides. Next, we consider the relative phase $\phi$
in Eq.~\eqref{eq:photonpair}. Finally, we describe the measures
taken to reduce the bandwidth of the photons.

\subsection{The waveguides}
Waveguides are used instead of bulk crystals because they yield a much
higher conversion efficiency. This is necessary because the spectral
filtering we apply is much narrower than the intrinsic spectral width
of the down-conversion process, so only a small fraction of the pump power
is used to create photons in the desired spectral range. Hence, the
larger conversion efficiency essentially compensates the loss in power
of the pump.

The photon pair source is based on two nonlinear waveguides made from
different materials and with different parameters. The choice of using two 
different types of waveguides was made for practical reasons that are not important for the results
presented in this paper. However, this choice allows for a direct comparison of
the performance of the two waveguides. A selection of parameters for the two waveguides is shown
in table Table~\ref{tab:waveguides}.

The first waveguide was obtained from AdvR~Inc.\ and has been fabricated
in a chip of periodically poled potassium titanyl phosphate (PPKTP)
by ion exchange. The chip contains a collection of
identical waveguides of width and height approximately \SI{4}{\micro
  m} and \SI{7}{\micro m}, respectively. Each waveguide spans the
entire \SI{13}{mm} length of the chip. The poling period of
\SI{8.2}{\micro m} allows to achieve type-I phase matching for the
signal and idler wavelengths of \SI{883}{nm} and \SI{1338}{nm} at a
temperature of about \SI{53}{\degreeCelsius}. The chip is heated to
this temperature using a custom oven based on a thermo-electric
cooler.  No dielectric coatings have been applied to the end faces of
the chip.  We previously used this waveguide, henceforth referred to
as the PPKTP waveguide, for the generation of narrowband photon pairs
in a series of experiments with solid-state quantum
memories~\cite{Clausen2011,Usmani2012,Clausen2012}.

The second waveguide was custom designed at the University of
Paderborn. It was fabricated by titanium indiffusion on a lithium
niobate chip. The chip is \SI{62}{mm} long and contains 25 groups of
\SI{50}{mm} long regions with poling periods between
\SI{6.40}{\micro m} and \SI{6.75}{\micro m}. Within each group there
are three waveguides of \SIlist{5;6;7}{\micro m} width,
respectively. We achieved the best results with a waveguide of poling
period \SI{6.45}{\micro m} and \SI{6}{\micro m} width, where the
temperature for type-I phase matching at the desired wavelengths is
about \SI{173}{\degreeCelsius}. The chip is heated to this temperature
with the help of an oven by Covesion~Ltd., which has been slightly
modified to accommodate the long chip. The elevated temperature is
chosen to mitigate the deterioation of the phasematching by photorefraction.

\begin{table}
  \caption{\label{tab:waveguides}A selection of the parameters of the
    two waveguides for direct comparison.}
  \begin{ruledtabular}
    \begin{tabular}{lcc} 
      & \multicolumn{2}{c}{Waveguide} \\ 
      & PPKTP & PPLN \\ [0.2em]
      \hline \\ [-.5em]
      Supplier & AdvR Inc. & Uni. Paderborn \\ 
      Poling period & \SI{8.2}{\micro m} & \SI{6.45}{\micro m} \\
      Length of poled region & \SI{13}{mm} & \SI{50}{mm} \\
      Waveguide width & $\sim$ \SI{4}{\micro m} & $\sim$ \SI{6}{\micro m} \\
      Waveguide height & $\sim$ \SI{7}{\micro m} & $\sim$ \SI{6}{\micro m} \\
      Phase-matching temperature & $\sim$ \SI{53}{\degreeCelsius} 
      & $\sim$ \SI{173}{\degreeCelsius} \\[0.2em]
    \end{tabular}
  \end{ruledtabular} 
\end{table}

The custom design of the second waveguide, from now on called the PPLN
waveguide, allowed for the addition of
a number of features which make it especially suitable for spontaneous
parametric down-conversion at the desired wavelengths. On the input
side, a $\lambda/4$ SiO$_2$-layer has been applied to the input face
to provide an anti-reflective coating for the pump laser at
\SI{532}{nm}. Additionally, the input side has a \SI{12}{mm} long
region without periodic poling where the waveguide width is linearly
increased from \SI{2}{\micro m} to the final width. Such a taper
should facilitate the coupling of the pump laser to the fundamental spatial
mode of the waveguide. The output side of the chip has been coated
with a 15-layer SiO$_2$/TiO$_2$ stack optimized for high reflection of
the pump light and high transmission of the signal and idler
photons. Measurements on a reference mirror that was coated
simultaneously with the chip revealed reflectivities of
\SIlist{94;2.4;12}{\percent} at \SIlist{532;880;1345}{nm},
respectively.

\subsection{Matching of the spatial modes}
To obtain a high degree of entanglement between the photon pairs
generated in the two waveguides, it is essential that the spatial
mode of the photon does not reveal in which waveguide it has been
created. A small mismatch can be corrected with a suitable
spatial-mode filter, such as a single-mode optical fiber. If, however,
the mismatch is large, the asymmetric losses introduced by the filter
can significantly reduce the amount of entanglement.

In theory, the use of identical waveguides should ensure a perfect
overlap of the spatial modes of the generated photons. In practice,
however, the production process often introduces small variations
between identically designed waveguides. In our case, the situation is
complicated by the fact that the waveguides are made of different
materials, have different dimensions and the signal and idler photons
are at widely separated wavelengths. In short, these factors make a
simple configuration with just a single interferometer, as depicted in
Fig.~\ref{fig:MachZehnder}, impossible for several reasons, in
particular when only a single aspheric lens is used to collect the
signal and idler photons at the output of the waveguides. Already for a single
waveguide, the chromatic aberration of the lense does not allow for
simultaneous collimation of the signal and idler beams. On top of that
there is the more fundamental problem that the refractive index
profiles of the waveguides depend on the chip and on the
wavelength. The result is that the signal and idler spatial modes have
different sizes and are not centered with respect to each other, even
if generated in the same waveguide. For
different waveguides, signal and idler beams can in general not be
pairwise matched by even the most sophisticated lens system.

\begin{figure}
  \centering
  \def\svgwidth{\columnwidth}
  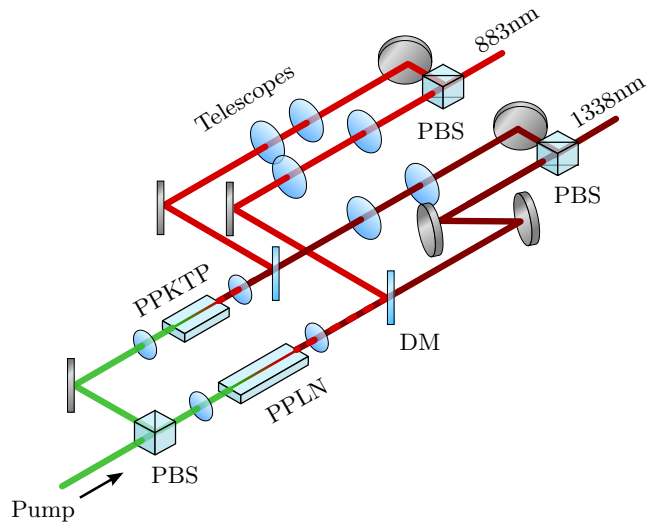
  \caption{The spatial modes of the photons generated in different
    waveguides can be efficiently matched by using two interleaved
    interferometers with appropriate telescopes.}
  \label{fig:dual-interferometer}
\end{figure}
One way to properly match the spatial modes is to part ways with
the idea of using a single interferometer and instead use two
interleaved interferometers, as shown in
Fig.~\ref{fig:dual-interferometer}. This gives control of all four
spatial modes involved. A single uncoated achromatic
lens (Thorlabs C220-TME) after each waveguide is positioned such that
the idler beams are collimated. Right after that, dichroic mirrors
separate signal and idler beams, leading to four individual beam
paths. Telescopes in three of the paths adapt the spatial modes such
that the signal and idler modes are separately matched to each
other and to the single-mode fibers that will eventually receive the
photons. Finally, the signal and idler modes are, respectively,
recombined on two PBSs. 

\subsection{Relative phase}
The relative phase from Eq.~\eqref{eq:photonpair} has contributions
from signal and idler photons, $\phi = \phi_s(\omega_s) +
\phi_i(\omega_i)$, and depends, in general, on the frequencies
$\omega_s$ and $\omega_i$ of the signal and idler photons,
respectively. In turn, $\phi_s$ is the
difference phase acquired between the horizontal and vertical paths of
the respective interferometer, and similarly for the idler photon. To
obtain a high degree of entanglement, it is important that $\phi$ is
well-defined for all frequencies within the final bandwidth of the
photons. Hence, the path length difference $\Delta L_x$ ($x=s,i$)
for the two interferometers should be much smaller than the coherence
length of the photons after spectral filtering. For the estimation
of $\Delta L_x$ one should not forget the dispersion inside the waveguides
and that also the propagation of the pump light up to the waveguides
is important.

In the experiment we actively stabilize $\phi$. For this purpose, each
interferometer contains a mirror mounted on a piezo-electric
transducer. We use the pump light at \SI{532}{nm} that is transmitted
through the waveguides and leaks into all parts of the interferometer
to continuously probe the phase. The PBSs at the input and outputs of
the interferometers are not perfect at this wavelength, such that 
residual interference can be seen on the intensity variations picked
up by two photodiodes. Note that, in general, the pump light
transmitted through the horizontal and vertical paths of the
interferometers will not have the same intensity. Additionally, the
coating on the end face of the PPLN chip, the reliance on
imperfections and the bad spatial mode-matching of the \SI{532}{nm}
light at the output result in peak-to-peak intensity variations as low
as a few ten nanowatts. Using a lock-in technique, an error signal can
nevertheless be extracted and used to stabilize the phases of the
interferometers.

Using this technique, the stabilization works reliably for a typical
duration of 5 to 10~hours, a duration after which the thermal drift in the
laboratory would typically exceed the compensation range of the piezos. However, the
technique has two limitations to keep in mind. First, the absolute
value of the phase can not be chosen at will and is more or less
random for every activation of the lock. Second, since the
\SI{532}{nm} light follows a slightly different path than the signal
and idler photons, and the temperature dependence of the refractive
index inside the waveguides is wavelength dependent, differential
phase shifts can appear. In practice, we observe residual phase drifts
on the order of \SI{1}{\degree/hour}, as determined by repeatedly
applying the measurement procedure described in Sec.~\ref{sec:entanglement}.

\subsection{Spectral filtering}
In experiments where one of the photons in a pair is coupled to a
narrowband receiver, such as an atomic ensemble, spectral filtering is
essential. In the typical scenario of SPDC with a narrowband
pump laser, energy conservation ensures that a detection of, say, the
idler photon after a suitable spectral filter guarantees that the
signal photon is within the target spectral range. At first glance
such one-sided filtering might seem entirely sufficient. In practice,
however, and in particular in the case of strong filtering, multi-pair
production can add a significant background of signal photons outside the
desired bandwidth, which leads to a reduction of the signal-to-noise
ratio of coincidence detections. Hence, also the signal photon needs
to be filtered at least to some extent.

Efficiency, stability and ease of use are typical criteria for
choosing suitable spectral filters. For a given bandwidth, one wants
to use as few filtering elements as possible, as all of them are bound
to introduce photon loss and have stabilization requirements. The case
of polarization-entangled photon pairs adds the concern that both the
spectrum and the efficiency of the filters need to be independent of
polarization. This precludes the use of traditional techniques such as
diffraction gratings, but also of some more recent developments such
as phase-shifted fiber Bragg gratings and Fabry-Perot cavities based
on coated lenses~\cite{Palittapongarnpim2012}. 

\begin{figure}
  \centering
  \includegraphics{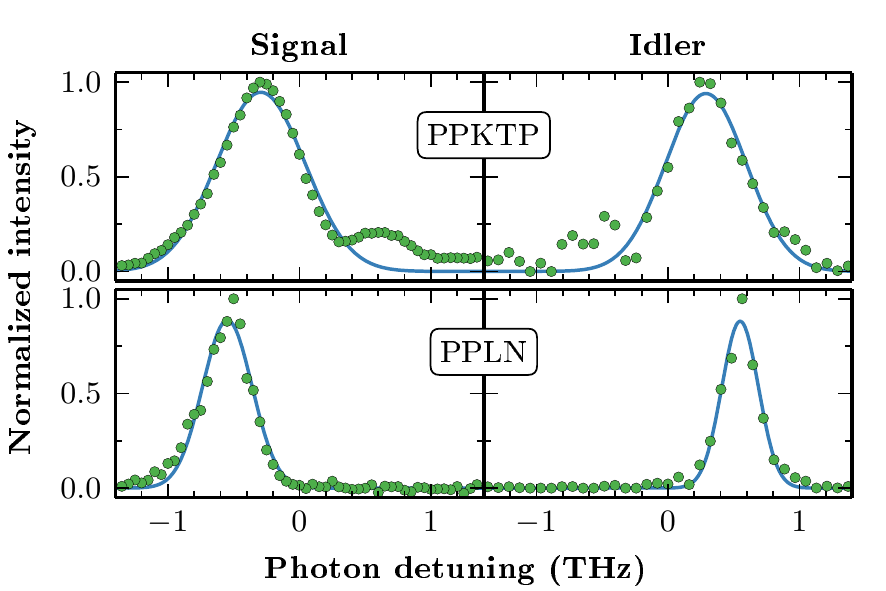}
  \caption{Non-filtered spectra of the photons generated by the two
    waveguides. Detunings are given with respect to a reference laser
    at \SI{883.2}{nm} for the signal photon, and for the idler with
    respect to light from difference-frequency generation using the
    same laser. Gaussian fits (solid lines) give estimates of the
    spectral bandwidths (see text). For these plots, the temperature
    of the waveguides had not yet been properly adjusted.}
  \label{fig:initial_spectra}
\end{figure}
The spectra of the two waveguides were measured using
custom-built spectrometers based on diffraction gratings and
single-photon-sensitive CCD cameras; see
Fig.~\ref{fig:initial_spectra}. The spectrometers have an estimated
resolution on the order of \SI{200}{GHz} FWHM at \SI{883}{nm} and \SI{100}{GHz}
at \SI{1338}{nm}. Gaussian fits to the
respective signal and idler spectra serve to estimate the
phasematching bandwidth. For the PPKTP waveguide the two fits
approximately agree, yielding a full width at half maximum (FWHM) of
\SI{791(28)}{GHz} for the signal and \SI{724(39)}{GHz} for the
idler. The signal photons generated in the PPLN waveguide are
measured to be \SI{443(12)}{GHz} wide, and the idler photons
\SI{328(11)}{GHz}. While both values may be resolution limited, the
discrepancy is most likely due to the inferior resolution at
\SI{883}{nm}.

Assuming the $\sinc^2$-shaped spectrum of ideal SPDC and neglecting
the dispersion caused by the refractive index profile of the waveguide, we can
use Sellmeier equations for KTP~\cite{Kato2002} and
LiNbO$_3$~\cite{Jundt1997} to find a theoretical
estimate of the bandwidths (see Appendix~\ref{app:pm-bandwidth}). For the waveguide from AdvR
the FWHM is estimated to \SI{540}{GHz}, while for the guide from
Paderborn we find \SI{100}{GHz}. In both cases, the measured
bandwidths are larger. Apart from the limited resolution of the
spectrometer, we attribute this deviation to inhomogeneities of the
waveguide structure over the interaction length, which also explains why the measured spectra do not
exhibit a $\sinc^2$ shape. Finally, propagation losses of the pump
laser in the waveguide can lead to a reduced effective interaction
length and hence a broadening of the spectra.

We shall now describe the filtering system used to reduce the spectral
width of the photon pairs to \SI{240}{MHz} FWHM\@. 
The filtering for the signal and idler photons is very similar and is
done in two steps. The signal photon is first sent onto a volume Bragg
grating (VBG) made by Optigrate. The VBG has a nominal diffraction
efficiency of \SI{98.6}{\percent}, although the value in the
experiment is  $\approx \SI{90}{\percent}$. The spectral selectivity is
specified to \SI{54}{GHz} at FWHM\@. Grating parameters are such that
the diffracted beam forms an angle of about \SI{7}{\degree} with the
incoming beam. We have not seen any polarization dependence of
significance in the performance of the VBG\@. The second filtering
step is an air-spaced Fabry-Perot etalon made by SLS Optics Ltd. The
etalon has a line width of $\Gamma_s/(2\pi) = \SI{600}{MHz}$ and a
free spectral range (FSR) of \SI{50}{GHz}, corresponding to a finesse
of 83. The peak transmission of the etalon is about \SI{80}{\percent}.

For the idler photon, the first filter is a custom-made Fabry-Perot
cavity with line width $\Gamma_i/(2\pi) = \SI{240}{MHz}$ and an FSR of
\SI{60}{GHz}, corresponding to a finesse of 250. By itself, we
achieved peak transmissions through the cavity exceeding
\SI{80}{\percent}. Integrated in the setup of the photon pair source,
mode matching was slightly worse, giving a typical transmission around
\SI{60}{\percent}. The cavity was followed by a VBG with a FWHM
diffraction window of \SI{27}{GHz} and nominal efficiency of
\SI{99.6}{\percent}. In this case, experimental observations were
compatible with specifications.

The idea behind the combination of Fabry-Perot filter and volume Bragg
grating is to select only a single longitudinal mode of the cavity or
the etalon. In practice, however, a typical reflection spectrum of a
VBG can have significant side lobes~\cite{Ciapurin2012}. From the
measured second-order auto-correlation functions (see
Sec.~\ref{sec:spectral-characterization}), we estimate that more than
\SI{70}{\percent} of the transmitted signal photons and more than
\SI{95}{\percent} of the idler photons belong to the desired
longitudinal mode.

One issue with narrowband filters is the spectral stability. Long-term
stability for the VBGs is easily achieved by using a stable mechanical
mount, as they have practically no sensitivity to temperature
fluctuations. The Fabry-Perot filters are stabilized in temperature,
but exhibit residual drifts on the order of \SI{100}{MHz/hour}. If the
center frequencies of the signal and idler filters drift such that
they no longer add up to the frequency of the pump laser, the
coincidence rate will drop. We compensate this by using a reference
laser at \SI{883}{nm}, which may be stabilized to the etalon, for
difference frequency generation (DFG) in the PPLN waveguide,
effectively giving coherent light at the idler frequency. The
frequency of the pump laser is then adjusted to optimize the
transmission of the DFG light through the cavity. During experiments,
we switch between DFG and SPDC every few tens of milliseconds, and the
transmitted DFG light is detected with single-photon detectors and
integrated over approximately \SI{1}{s}. The stabilization was
implemented in software for previous
work~\cite{Clausen2011,Usmani2012,Clausen2012}, and reliably
compensates the slow and weak thermal drifts.

\section{Spectral characterization via correlation functions}
\label{sec:spectral-characterization}
Correlation functions are a useful tool for the characterization of
light sources. We consider, in particular, the normalized second-order correlation
functions, which are unaffected by photon loss or detector
inefficiency. They are defined as
\begin{equation}
  \label{eq:g2-definition}
  g_{jk}^{(2)}(\tau) \equiv \frac{\expect{E_j^\dag(t) E_k^\dag(t+\tau)
      E_k(t+\tau) E_j(t)}}{\expect{E_j^\dag(t)
      E_j(t)}\expect{E_k^\dag(t+\tau) E_k(t+\tau)}},
\end{equation}
where the indices $j,k \in \{s,i\}$ represent the signal or idler
photon, respectively. A measurement of $g_{jk}^{(2)}(\tau)$ consists
of first determining the rate of coincidence detections between modes
$j$ and $k$ at a time delay $\tau$. This is effectively a measurement
of the non-normalized second-order coherence function, which is the
numerator in Eq.~\eqref{eq:g2-definition}. The normalization is then
performed with respect to the rate of coincidences between photons from
uncorrelated pairs
created at times differing by much more than the coherence
time of the photons. 

By itself, the second-order cross-correlation function
$g_{si}^{(2)}(\tau)$ gives a measure of the quality of a photon-pair
source, because noise photons stemming from imperfect spectral
filtering or fluorescence generated in the down-conversion crystal
inevidently reduce the amount of correlations. The auto-correlation
functions $g_{ss}^{(2)}(\tau)$ and $g_{ii}^{(2)}(\tau)$ give
information about the multimode character of the photons and their spectra. Finally, the
cross- and auto-correlation functions can be combined in a
Cauchy-Schwarz inequality whose violation proves the quantum character
of the photon-pair source~\cite{Kuzmich2003}.

In this section we look at the normalized auto- and cross-correlation
functions of the signal and idler photons. We show that the shape of
the correlation functions is exactly as one would expect from the
spectral filtering, if the jitter of the detectors is taken properly
into account. Additionally, we use the auto-correlation functions to
deduce the probability that a detected signal (or idler) photon stems
from the desired mode of the filtering etalon (or cavity).

\subsection{Correlation functions}
\label{sec:correlation-functions}
The spectral filtering reduces the uncertainty in energy of the signal
and idler photons. The effect can be directly seen on the normalized
second-order auto- and cross-correlation functions, for which simple
analytical expressions can be derived for collinear, low-gain, SPDC
with plane-wave fields. The detailed derivation is given in
Appendix~\ref{app:pairsource-theory}. In brief, it procedes as
follows.  First, expressions for the first-order field correlation
functions without filtering can be obtained via the Bogliubov
transformation that describes the input-output relation of the SPDC
process~\cite{Razavi2009,Wong2006}. Next, spectral filtering is
included through the convolution of the correlation functions with the
filter impulse response~\cite{Mitchell2009}. In the case where the
bandwidth of the filters is much smaller than the bandwidth of the
SPDC process, the temporal dependence of the correlation functions is
entirely given by the spectral filtering. Finally, higher-order
correlation functions are obtained by applying the quantum form of the
Gaussian moment-factoring theorem~\cite{Razavi2009}. We arrive at the
following expressions for the normalized second-order cross- and
auto-correlation functions for Lorentzian-shaped spectral filters,
\begin{equation}
  \label{eq:g2-summary}
  \begin{aligned}
    g_{si}^{(2)}(\tau) &= 1 + 4\frac{B}{R} \frac{\Gamma_s
      \Gamma_i}{(\Gamma_s + \Gamma_i)^2}\, f_{si}(\tau) \\
    g_{ss}^{(2)}(\tau) &= 1 + f_{ss}(\tau) \\
    g_{ii}^{(2)}(\tau) &= 1 + f_{ii}(\tau),
  \end{aligned}
\end{equation}
where the temporal dependence is given by
\begin{equation}
  \label{eq:g2-temporal-dependence}
  f_{jk}(\tau) =
  \begin{cases}
    e^{\Gamma_j \tau} & \text{for } \tau < 0 \\
    e^{-\Gamma_k \tau} & \text{for } \tau \geq 0
  \end{cases}.
\end{equation}
The cross-correlation function depends on the inverse of the ratio of
the $R/B$. Here, $B$ is the phase-matching bandwidth and $R$ is the
rate of photon pair creation. Hence, $1/B$ is seen as the duration of one temporal mode. 
The low-gain limit of the source is obtained with the probability to create a pair per temporal mode
is much smaller than one, i.e.~$R/B \ll 1$. In this regime, the rate $R$ is proportional
to the pump power. Additionally, the
cross-correlation depends on the ratio of the filter bandwidths. For a
given value of $R/B$, a larger mismatch makes it more likely that only
one of the photons in a pair passes the filters, which leads to a
reduction of the cross-correlation.

\subsection{Detector jitter}
\label{sec:correlations-jitter}

\begin{figure}
  \centering
  \includegraphics{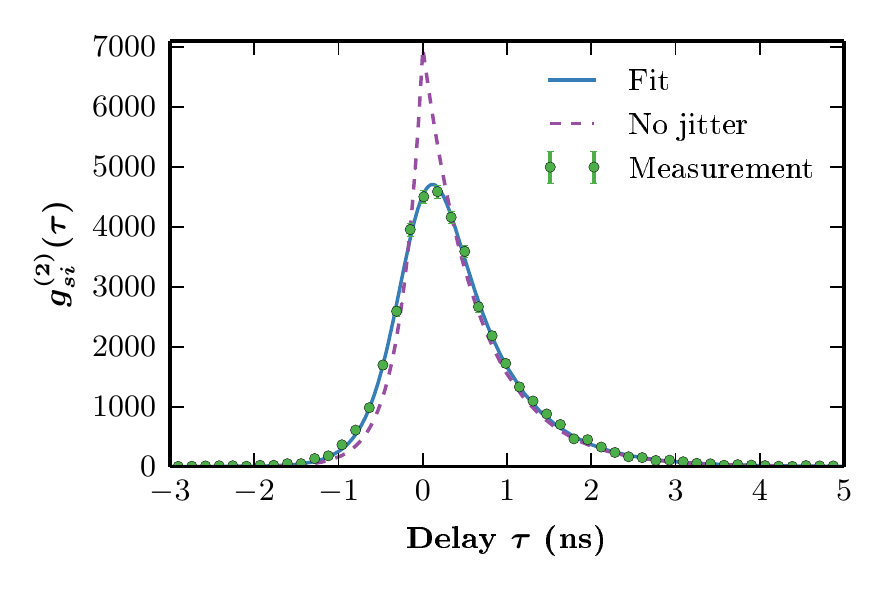}
  \caption{Example of a cross-correlation function measured for the
    PPKTP waveguide using a binning of \SI{162}{ps}. The solid line is
    a fit to the theoretical line shape (Eq.~\ref{eq:g2-summary}),
    corrected for detector jitter, where the only free parameters are
    the ratio $R/B$ and a horizontal offset. The dashed line is the
    cross-correlation that we could have obtained with a jitter-free
    detection system.}
  \label{fig:cross-correlation}
\end{figure}

Figure~\ref{fig:cross-correlation} shows an example of a measured
cross-correlation function for the PPKTP waveguide. The combination of
detectors, a Perkin-Elmer SPCM-AQRH-13 silicon avalanche photo diode
and a super-conducting nanowire single-photon detector (SNSPD), had
negligible dark count rates. To compare the measured temporal
dependence with theory, the jitter of the detection system has to be
taken into account. This can be done by convoluting the expression in
Eq.~\eqref{eq:g2-temporal-dependence} with the distribution function
of the jitter. In our case the jitter is well modeled by a normal
distribution, and the expression for the refined temporal dependence
$\tilde{f}_{jk}(\tau)$ is given in the appendix. After this
modification, we find excellent agreement between the measurement and
a theoretical fit, where the only free parameters are a horizontal
offset and the ratio $R/B$. Note that the jitter of
$\sigma=\SI{250}{ps}$ for this combination of detectors reduces the
maximum cross-correlation by a factor $\tilde{f}_{si}(0) = 0.65$.

\subsection{Multimode properties}
\label{sec:correlations-multimode}
Contrary to the cross-correlation function, the normalized
auto-correlation functions do not depend on the spectral
brightness. Instead, they reach a maximum value of
$g_{jj}^{(2)}(0) = 2$, which reveals the thermal nature of the
individual signal and idler fields.

\begin{figure}
  \centering
  \includegraphics{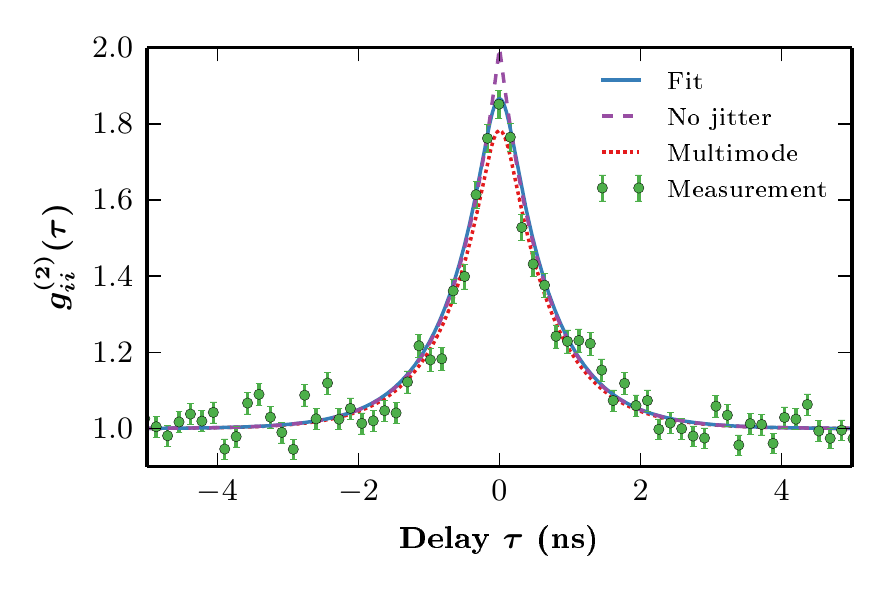}
  \caption{The second-order auto-correlation function of the idler
    photons generated in the PPKTP waveguide. Bins are
    \SI{162}{ps}. The solid line is a fit to the theoretical
    line shape (Eq.~\ref{eq:g2-summary} with jitter included), where the
    only free parameter is a horizontal offset. The dashed line is the
    auto-correlation that we could have obtained with a jitter-free
    detection system. The dotted line is a simulation
    corresponding to a \SI{2.5}{\percent} occupation of each 
    nearest-neighbor longitudinal cavity mode.}
  \label{fig:g2idler}
\end{figure}

A comparison between theory and experiment for the auto-correlation
function of the idler photons generated in the PPKTP waveguide is
plotted in Fig.~\ref{fig:g2idler}. Detector jitter has been included
as before by using $\tilde{f}_{ii}(\tau)$ instead of
$f_{ii}(\tau)$. The detectors were a pair of SNSPDs with $\sigma =
\SI{125}{ps}$. The theoretical prediction is in excellent agreement
with the measured data.

A measurement of the second-order auto-correlation function allows,
additionally, to characterize the presence of spurios spectral modes,
that is, undesired modes of the Fabry-Perot filters, in the signal and
idler fields. This has first been shown for pulsed and broadband SPDC
in~\cite{Christ2011}, where a set of orthogonal spectral modes is
obtained via Schmidt decomposition of the joint-spectral amplitude
of the signal and idler fields. By normalizing the occupation
probabilities $p_n$ of these modes such that $\sum p_n = 1$, the
authors define an effective number of modes $K = 1/\sum p_n^2$. This
number, also known as the Schmidt number, quantifies the amount of
spectral entanglement and is the reciprocal of the purity of the
reduced states of the signal and idler
modes~\cite{Eberly2006}. Furthermore, it is shown in~\cite{Christ2011}
that the inability to resolve these spectral modes results in a
reduction of the auto-correlation functions, given by $g_{jj}^{(2)}(0)
= 1 + \sfrac{1}{K}$. Hence, a measurement of $g_{jj}^{(2)}(0)$ allows
to directly determine $K$.

For continuous-wave SPDC subjected to narrow-band Fabry-Perot filters,
the longitudinal modes of the filter form a suitable basis for the
spectral decomposition. We define $p_0$ as the probability to find the
photon in the desired longitudinal mode, and let $p_n$ be the $n$-th
red-detuned (or blue-detuned) mode for $n>0$ (or $n<0$). We would like
to determine a lower bound on $p_0$ via a measurement of the
auto-correlation function. As in the case of pulsed SPDC, the presence
of spurious longitudinal modes of the Fabry-Perot filter reduces the
auto-correlation function. This is easily seen from the fact that
$f_{jj}(\tau)$ is proportional to the absolute square of the Fourier
transform of the power spectral density of the cavity transfer
function (see also Eqs.~\eqref{eq:lorentzian-transfer}
and~\eqref{eq:auto-correlation-final}). The presence of multiple
longitudinal cavity modes will hence lead to oscillations of
$g_{jj}^{(2)}(\tau)$ at a frequency corresponding to the free spectral
range of the filter. If the detectors do not resolve these
oscillations, they will be averaged out, leading to a reduction of
$g_{jj}^{(2)}(\tau)$. However, in our case the detector jitter is
sufficiently strong to give a reduction of the $g_{jj}^{(2)}(0)$ even
for the single-mode case. To more clearly separate the contributions
from detector jitter and spurious modes, we rewrite the
auto-correlation function of Eq.~\eqref{eq:g2-definition} as
\begin{equation}
  \label{eq:g2-multimode}
  g_{jj}^{(2)}(\tau) = 1 + \frac{1}{K}\ \tilde{f}_{jj}(\tau),
\end{equation}
where jitter has been taken into account explicitely via the use of
$\tilde{f}_{jj}(\tau)$.

\begin{figure}
  \centering
  \includegraphics{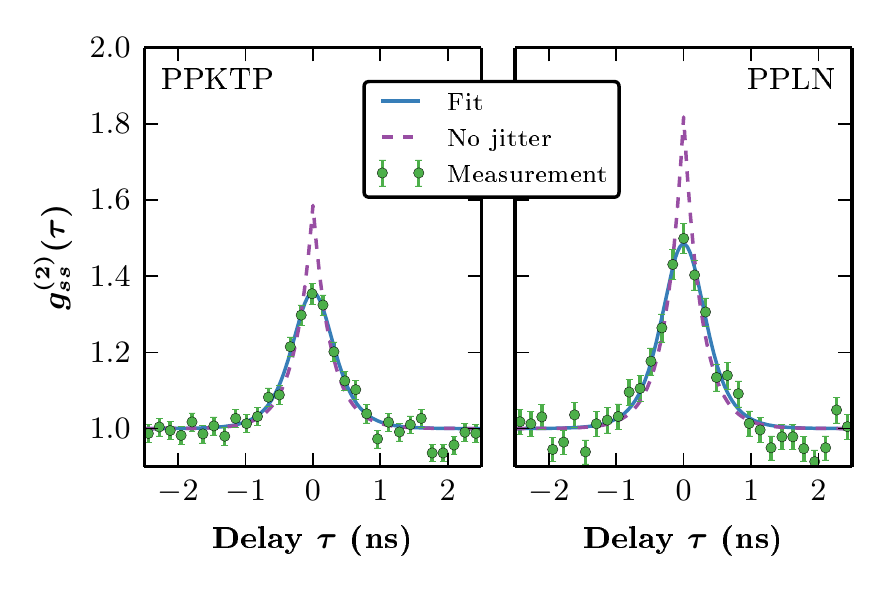}
  \caption{The second-order auto-correlation function of the signal
    photons generated in the PPKTP (left) and PPLN (right)
    waveguides. Spurious etalon modes prevent the peak to reach a
    value of 2, even after the correction for detector jitter. Bins
    are \SI{162}{ps}}
  \label{fig:g2signal}
\end{figure}

For the idler photon, the red dotted line in Fig.~\ref{fig:g2idler}
shows the case of $p_0=0.95$ for the central cavity mode and $p_{\pm
  1}=0.025$ for the neighboring red- or blue-detuned modes, giving
$K=1.1$. The mismatch with the experimental data at zero delay
is consistent with the selection of a single cavity mode by the
filtering system.

The situation is different for signal photon, for which
auto-correlation measurements are shown in
Fig.~\ref{fig:g2signal}. Here, the bandwidth of the volume Bragg
grating is comparable to the free spectral range of the etalon, and
contributions from spurious modes are to be expected. From a fit of
Eq.~\eqref{eq:g2-multimode} to the data, with $K$ and $\sigma$ as free
parameters, we obtain $K=\num{1.71(8)}$ for the PPKTP
waveguide and $K=\num{1.22(6)}$ for the PPLN waveguide. Assuming the worst
case of only a total of two etalon modes with non-zero population,
this corresponds to probabilities of $p_0 = \num{0.71(3)}$ and $p_0 =
\num{0.90(3)}$, respectively, for the photon being in the desired
etalon mode. We attribute the larger value of $K$ for the PPKTP
waveguide to the larger phase-matching bandwidth.

\section{Efficiency characterization of the filtered photon sources}
\label{sec:characterization}
\begin{figure*}
  \centering
  \includegraphics{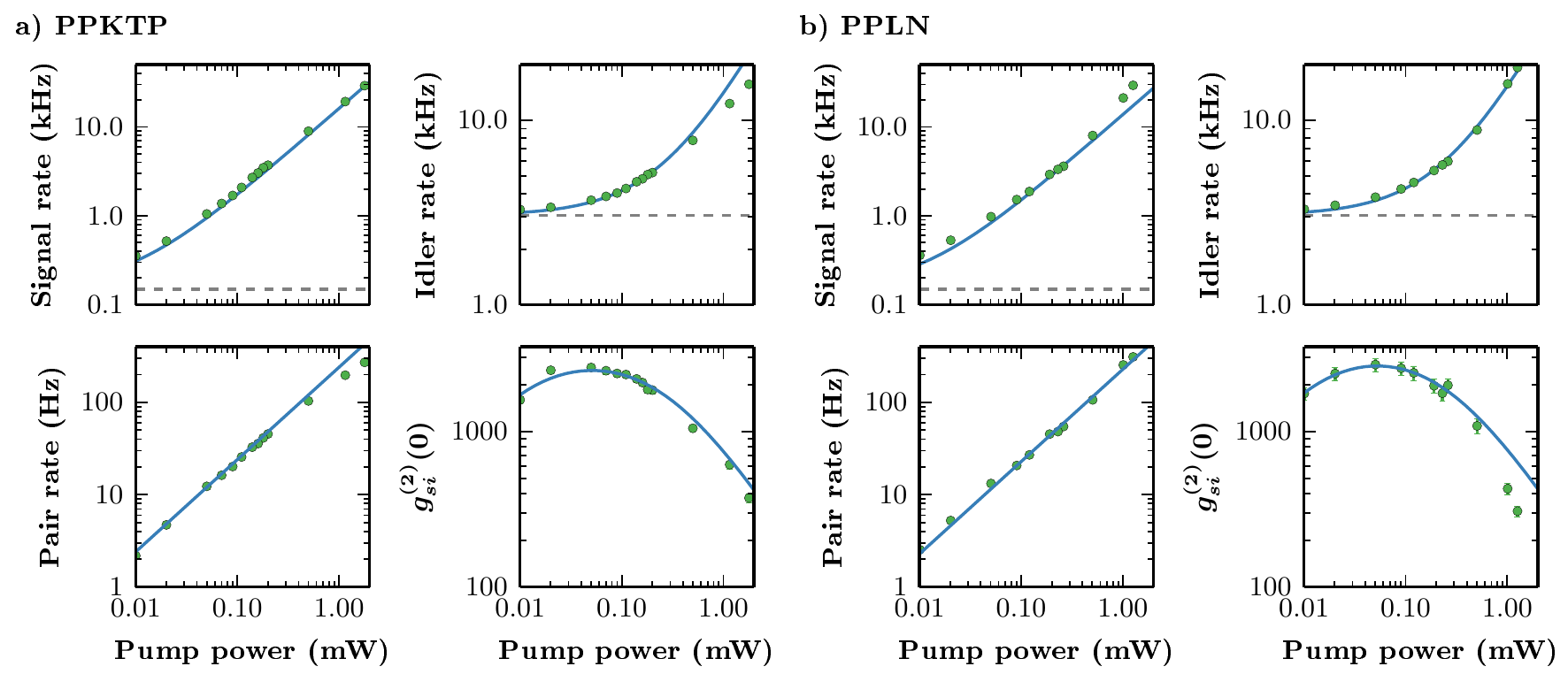}
  \caption{Characterization of a) the PPKTP and b) the PPLN waveguide.
    For each waveguide, the signal, idler and pair
    detection rates are plotted, as well as the value of the
    cross-correlation function at $\tau=0$ delay. The dashed
    horizontal lines in the panels for the signal and idler rates
    indicate the detector noise level. For the measurement
    of the pair rate, a coincidence window of \SI{6}{ns} was used,
    which is sufficiently large to encompass the entire coincidence
    peak (see Fig~\ref{fig:cross-correlation}). Additionally,
    accidental coincidences have been subtracted. The values of the
    cross-correlation function are based on a binning of
    \SI{162}{ps}. A common fit (solid lines) to all four data sets for
    each waveguide was used to extract the spectral brightness and
    collection efficiencies (see also Table~\ref{tab:characterization}).}
  \label{fig:characterization}
\end{figure*}
In this section we show a characterization of the individual
performances of the two waveguides, including spectral filtering. The
characterization aims at determining the spectral brightness and the
collection and detection efficiencies of the photons. It consists of
measuring as a function of the pump power the detection rates of
signal and idler photons. Furthermore, we measured the photon-pair
rate, that is, the signal-idler coincidence rate, corrected for
accidental coincidences, for a coincidence window that is large
compared to the coherence time. Finally, we also determined the power-dependence
of the second-order cross-correlation function $g_{si}^{(2)}(\tau)$ at
delay $\tau = 0$. The results are shown in
Fig.~\ref{fig:characterization}.

For comparison to a theoretical model, we use the same derivation as
for the correlation functions in the previous section. However, in the
previous section the dark counts of the detectors were
negligible. Dark counts add an offset to the signal and idler
detection rates. Additionally, they give rise to accidental
coincidences, which set an upper bound on the normalized
cross-correlation function at low pump powers. We included the dark
count rate $D_j$ in the model and also added finite detection efficiencies
$\eta_j$ to end up with the following set of equations (see also
Appendix~\ref{app:pairsource-theory}),
\begin{equation}
  \label{eq:characterization-fit}
  \begin{split}
    W_s &= \frac{1}{4}\frac{\eta_s}{p_0}\frac{R}{B} \Gamma_s + D_s \\
    W_i &= \frac{1}{4}\eta_i \frac{R}{B} \Gamma_i + D_i \\
    W_2 & = \frac{1}{4} \eta_s \eta_i \frac{R}{B} \frac{\Gamma_s
      \Gamma_i}{\Gamma_s + \Gamma_i} \\
    g_{si}^{(2)}(0) &= 1 + \frac{1}{4} \frac{\tilde{f}_{si}(0)\, \eta_s \eta_i}{W_s
      W_i} \frac{R}{B} \left( \frac{\Gamma_s \Gamma_i}{\Gamma_s +
        \Gamma_i} \right)^2.
  \end{split}
\end{equation}
Here, the signal and idler rates $W_s$ and $W_i$ are essentially given
by the spectral brightness of the waveguide times the respective
bandwidth of the filtering system and attenuated by the detection
efficiency. Since $R$ is proportional to the pump power, so are $W_s$
and $W_i$. $W_s$ has also been corrected for the contribution of
spurious etalon modes, which will increase the detection rate by a
factor $1/p_0$. The behavior of the pair rate $W_2$ is similar, except
that the photon pairs have an effective bandwidth of $\Gamma_s
\Gamma_i / (\Gamma_s + \Gamma_i)$, which is smaller than the bandwidth
of the signal and idler photons individually. Note that the
measurement of $W_2$ includes correction for accidental coincidences,
and no correction for dark counts needs to be applied to the
theory. Finally, the expression for $g_{si}^{(2)}(0)$ is equivalent to
the one given in Eq.~\eqref{eq:g2-summary}, but the inclusion of dark
counts prevents further simplification.

We used commercially available detectors for the measurements
presented in Fig.~\ref{fig:characterization}. The signal detector
by Perkin-Elmer has  dark-count rate of \SI{150}{Hz}
and a detection efficiency of about \SI{30}{\percent} at
\SI{880}{nm}. As detector for the idler photon served an ID220 by Id
Quantique with \SI{20}{\percent} efficiency. To reduce the contribution of
afterpulsing, the dead time of this detector was set to \SI{20}{\micro
s}, and we observed a dark-count rate of \SI{3.0}{kHz}. The
offset on the signal and idler count rates given by the dark counts is
indicated by dashed lines in the top panels of
Fig.~\ref{fig:characterization}.

\begin{table}
  \caption{\label{tab:characterization} Parameters as extracted from
    fitting the data in Fig.~\ref{fig:characterization} to
    Eqs.~\eqref{eq:characterization-fit}. $2\pi\, R/B$ is the spectral
    brightness, given in conventional units, for a pump power of~\SI{1}{mW}. $\eta_s$ (or $\eta_i$) is the
    overall collection and detection efficiency for the signal (or
    idler) photon.
  }
  \begin{ruledtabular}
    \begin{tabular}{ccc} 
      & \multicolumn{2}{c}{Waveguide} \\ 
      Parameter & PPKTP & PPLN \\ [0.2em]
      \hline \\ [-.5em]
      $2\pi\, R/B$ & \SI{2.45(6)e3}{\per(s MHz)} & \SI{3.08(6)e3}{\per(s MHz)} \\
      $\eta_s$ & \SI{3.1(2)}{\percent} & \SI{2.6(2)}{\percent} \\
      $\eta_i$ & \SI{7.4(1)}{\percent} & \SI{6.6(1)}{\percent} \\
    \end{tabular}
  \end{ruledtabular} 
\end{table}

A simultaneous fit to the Eqs.~\eqref{eq:characterization-fit}
reproduces the measurements to a high extent. The free parameters in
the fit are the spectral brightness $R/B$ and the overall collection
and detection efficiencies $\eta_s$ and $\eta_i$. The results of the
fit are shown in Table~\ref{tab:characterization}. For the PPKTP
waveguide the idler rate shows a negative deviation from the expected
behavior at pump powers above \SI{1}{mW}, where the detector starts
being saturated. For the PPLN waveguide the saturation seems to be
compensated by a higher pair-creation efficiency, indicated by a
positive deviation of the signal rate and a significant drop in the
cross-correlation. 

In terms of the spectral brightness, the two waveguides perform on a
similar level. We note however, that the specified pump power is
measured in front of the waveguide. For both waveguides we estimate a
total coupling of the pump laser into the waveguide is between
\SI{40}{\percent} and \SI{50}{\percent}. Of this, only a fraction is
coupled into the fundamental spatial mode, and hence contributing to
SPDC\@.  In principle, we would expect a higher brightness for the
waveguide from Paderborn, since it is longer and PPLN has a larger
non-linear coefficient than PPKTP\@. The reason that we observe
something different could be a non-optimal temperature of this
waveguide in this measurement, which shifts the perfect phase matching
slightly away from the filter transmission maximum. We also note that
at pump powers above a few milliwatts, the operation of the PPLN
waveguide is impaired by photorefraction, which leads to strong
fluctuations of the spatial mode of the pump laser inside the
waveguide.

In our experiments we are rarely constrained by the available pump
laser power, and the spectral brightness is only of minor
importance. More important are the achievable coincidence rates and
the correlations between signal and idler photons. The coincidence
rate is proportional to the product of the signal and idler collection
and detection efficiencies, $\eta_s$ and $\eta_i$. Also here we see
similar values for the two waveguides, indicating a spatial
mode-matching better than \SI{80}{\percent} for the
signal photon and around \SI{90}{\percent} for the idler. The
expected peak transmission for the signal path is $\eta_s
\approx \SI{3.6}{\percent}$ with contributions from a long-pass filter
that removes the pump light (\SI{80}{\percent}), the VBG
(\SI{90}{\percent}), the etalon (\SI{80}{\percent}), fiber coupling
(\SI{60}{\percent}) and detector efficiency
(\SI{30}{\percent}). Additionally, the setup was already prepared for
storage and retrieval in the quantum memory, adding losses due to a
fiber-optical switch (\SI{70}{\%}), fiber connectors (\SI{70}{\%}) and
another fiber coupling (\SI{70}{\%}). On the idler side, we expect
$\eta_i \approx \SI{8}{\percent}$, distributed over the cavity
(\SI{60}{\percent}), fiber coupling (\SI{70}{\percent}) and detector
efficiency (\SI{20}{\percent}). The measured value for $\eta_s$ and
$\eta_i$, given in Table~\ref{tab:characterization}, corresponds quite
well to the expected values. We attribute the small differences to
loss inside and at the end facets of the waveguides.

The measured cross-correlation function reaches for both waveguides a
peak value of approximately 2600 at a pump power of~\SI{50}{\micro
 W}. At lower pump power correlations are reduced by dark counts, at
higher pump powers by multi-pair emission.

\section{Entanglement}
\label{sec:entanglement}
The characterization of the two waveguides showed that a very high
degree of mode-matching for the photons originating from the two
waveguides has been obtained. Additionally, the spectral brightness is
about the same. This means that it should be possible to achieve a
high degree of entanglement by setting the pump polarization to
an approximately equal superposition of horizontal and vertical, such
that similar amounts of light arrive at the two waveguides. In
practive, we neglect the small differences in coupling efficiencies
and adjust the pump polarization such that the rate of coincidences
from the two waveguides is about the same. It remains to be shown that
the horizontally and vertically polarized photon pairs form a coherent
superposition with a stable phase, which corresponds to an entangled
state between the two photons. 

\begin{figure}
  \begin{tikzpicture}
    \node [anchor=north west] at (0,0) {\textbf{a)}};
    \node [anchor=north west] at (0,0)
      {\includegraphics{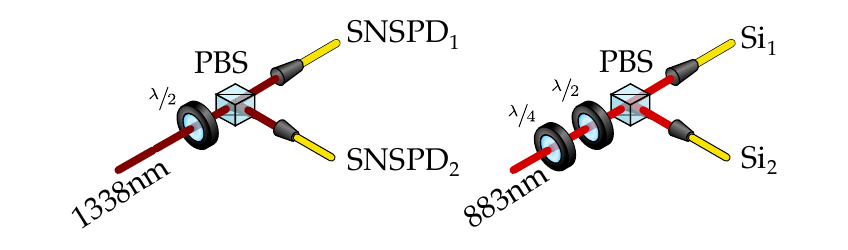}};
    \node [anchor=north west] at (0,-2.5)
      {\includegraphics{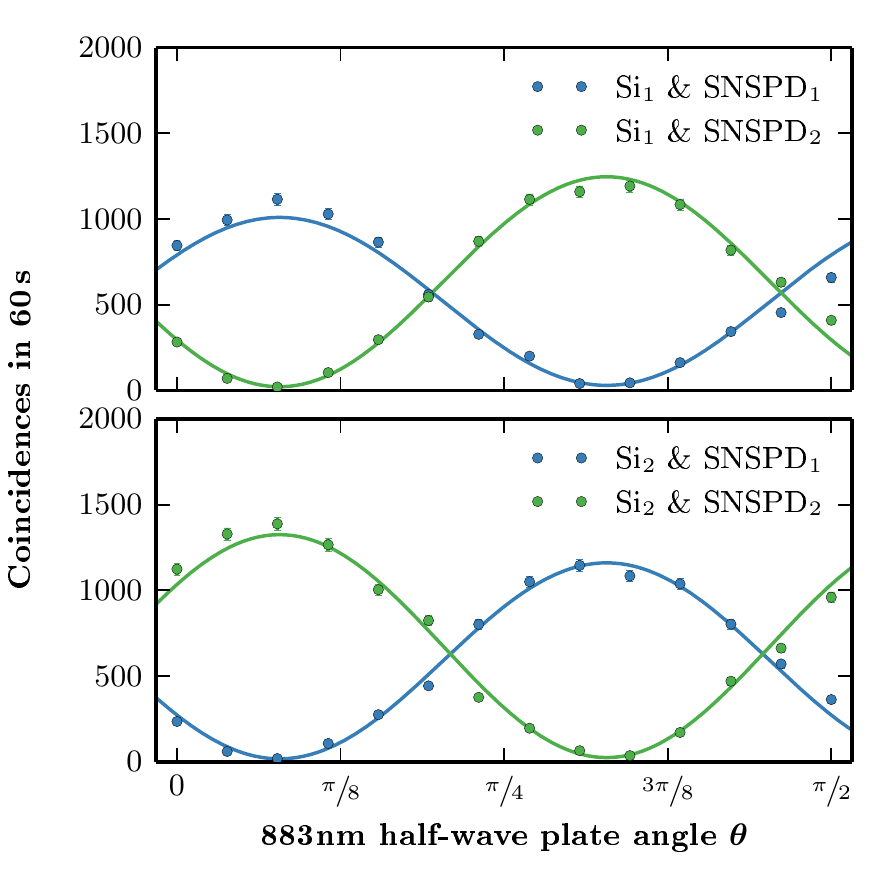}};
    \node [anchor=north west] at (0,-2.5) {\textbf{b)}};
  \end{tikzpicture}
  \caption{Characterization of the coherence of the pair source. a)
    The idler photons are measured in the bases of diagonal
    polarization. This projects the signal photon onto a coherent
    superposition of $\ket{H}$ and $\ket{V}$ with unknown relative
    phase. A quarter-wave plate at fixed angle transforms this state
    into a linear polarization, which is analyzed with the help of a
    half-wave plate and a polarizing beam splitter. b) Corresponding
    coincidence measurement for a coincidence window of \SI{2}{ns} for
    the four detector combinations. Solid lines are sinusoidal fits
    with a fixed period and common phase. The fits yield an average
    visibility of $V=\SI{96.1(9)}{\percent}$.}
  \label{fig:source-visibility}
\end{figure}

Let us, for simplicity, assume that the photon pairs are produced in the
maximally entangled state
\begin{equation}
  \label{eq:max-entangled-pair}
  \frac{1}{\sqrt{2}}\left(\ket{HH} +
    e^{i\phi}\ket{VV}\right).
\end{equation}
A measurement that verifies the coherent nature of this state is
illustrated in Fig.~\ref{fig:source-visibility}a. First, the idler
photon is measured in the basis $\ket{\pm} = (\ket{H} \pm
\ket{V})/\sqrt{2}$ using a half-wave plate and a polarizing beam
splitter. If a photon is detected in the port of the beam splitter
corresponding to, say, $\ket{+}$, the signal photon is projected onto
the state $\ket{\phi_+} = (\ket{H} + e^{i\phi}
\ket{V})/\sqrt{2}$. Sending this through a quarter-wave plate and a
half-wave plate whose fast axes are at angles of $\pi/4$ and $\theta$
to horizontal, respectively, transforms the signal photon into the
linearly polarized state $\ket{\beta} = \sin \beta \ket{H} - \cos
\beta \ket{V}$ with $\beta = 2\theta + \phi/2 + \pi/4$. We hence
expect that the probability of detecting the signal photon after a
polarizing beam splitter shows sinusoidal fringes as a function of
$\theta$ with a period of $\pi/2$. The phase of the fringes depends on
the phase $\phi$ of the initial entangled state
\eqref{eq:max-entangled-pair}, such that this kind of measurement can
be used to determine $\phi$. If, instead, the photon pairs are
generated in a maximally mixed state
$(\ketbra{HH}{HH}+\ketbra{VV}{VV})/2$, the same measurement of the
coincidence rate will not show any dependence on $\theta$. A
fringe visibility larger than \SI{33}{\percent} is necessary to infer
the presence of entanglement~\cite{Peres1996}.

In Fig.~\ref{fig:source-visibility}b we show the outcome of the
described measurement procedure. A pair of super-conducting nano-wire
single-photon detectors (SNSPDs) has been used for the idler photon,
and Si avalanche photo diodes (Perkin-Elmer) for the signal
photon. For each value of $\theta$ the number of coincidences in a
\SI{2}{ns} window have been integrated over a duration of
\SI{60}{seconds} for each of the four possibly detector
combinations. The number of measured coincidences oscillates as a
function of $\theta$, as expected. A sinusoidal fit reveals an average
visibility $V=\SI{96.1(9)}{\percent}$, which indicates that the source
generates photon pairs that are close to maximally entangled in
polarization.

To unequivocally prove the presence of entanglement we performed a
violation of the Clauser-Horne-Shimony-Holt (CHSH)
inequality~\cite{Clauser1969}. A quarter-wave plate was added to the
polarization analysis of the idler photon, such that the setups for
signal and idler photon of Fig.~\ref{fig:source-visibility}a were now
identical. Additionally, the SNSPDs were replaced by ID220s for their
higher detection efficiency. The wave plate allows to switch the
measurement basis of the idler photon between $\ket{\pm}$ and the
circular polarizations $(\ket{H} \pm i \ket{V})/\sqrt{2}$ by a
rotation of the half-wave plate. These two basis sets were used for
the measurement. Since we do not \emph{a priori} know the relative
phase $\phi$ of the photon pairs, we determine the optimal settings
for the signal analyzer as follows. We set the idler analyzer to
$\ket{\pm}$ and perform another measurement of the type of
Fig.~\ref{fig:source-visibility} to determine the angle
$\theta_\text{max}$ of the half-wave plate of the signal analyzer that
gives a maximum between detectors Si$_1$ and ID220$_1$. For the
violation of the CHSH inequality we then use the angles
$\theta_\text{max} \pm \sfrac{\pi}{16}$. For an acquisition time of
\SI{5}{minutes} per setting we find a CHSH parameter of
$S=2.708(9)$, which is almost 80 standard deviations above the
bound for separable states of $S\leq 2$.

\section{Summary and Outlook}
\label{sec:outlook}
We have presented a source of polarization-entangled photon pairs
based on the nonlinear waveguides of different materials embedded in
the arms of a polarization interferometer. We have shown
that the source emits photon pairs with a high degree of entanglement
and is compatible with the storage of one of the photons in a quantum
memory. The wavelength of the other photon is in a telecom window,
which permits the low-loss transmission over optical fiber. This
combination makes the source particularly useful for 
quantum communication experiments.

Even though the photon-pair source is conceptually simple, a higher
degree of integration would be desirable. Recent work along this
direction includes the integrated spatial separation of signal and
idler photons using an on-chip wavelength-division
multiplexer~\cite{krapick2013} and the direct generation of
\SI{150}{MHz} broad photon pairs using a monolithic waveguide
resonator~\cite{Luo2013a}. Both of these techniques were demonstrated
with similar wavelengths as used in this work. In particular the
latter could greatly simplify the efficient generation of narrowband
photon pairs, provided that the intrinsic resonator loss can be
reduced. If this could further be combined with the on-chip generation
of polarization-entangled photons using an interlaced bi-periodic
structure~\cite{Herrmann2013}, one would have the equivalent of the
whole setup of Fig.~\ref{fig:dual-interferometer} on a single chip,
including spectral filtering. Together with the recent progress in
solid-state quantum memories, these are promising perspectives for the
development of compact and practical nodes for quantum communication.

\begin{acknowledgments}
  This work was supported by the Swiss NCCR
  Quantum Science Technology as well as by the European project
  QuRep. We thank Rob Thew, Anthony Martin, Hugues de Riedmatten and Jonathan Lavoie for useful discussions.
\end{acknowledgments}

\appendix

\section{Estimation of phase-matching bandwidth}
\label{app:pm-bandwidth}
The frequency dependence of spontaneous parametric down-conversion is
given by the joint spectral amplitude $f(\omega_s, \omega_i)$, which
can be written as the product of two functions,
\begin{equation}
  \label{eq:jsa}
  f(\omega_s, \omega_i) = \alpha(\omega_s + \omega_i)\, \Phi(\omega_s,
  \omega_i),
\end{equation}
where $\omega_s$ (or $\omega_i$) is the frequency of the signal (or
idler) photon, $\alpha(\omega)$ represents the spectrum of the pump laser and
$\Phi(\omega_s, \omega_i) = \sinc(\Delta k L/2)$
is the phase-matching function. The state of a single
photon pair can be written in terms of the joint spectral amplitude as
\begin{equation}
  \label{eq:spectral-wavefunction}
  \ket{\Psi} \propto \infint[d\omega_s]
  \infint[d\omega_i] f(\omega_s, \omega_i)\,
  a^\dag(\omega_s)\, a^\dag(\omega_i)\ket{\text{vac}},
\end{equation}
where $a^\dag(\omega)$ is the photon creation operator at
frequency~$\omega$. We recognize, that $f(\omega_s, \omega_i)$ is the
spectral wavefunction of the photon pair. It follows that the spectral
distribution, that is, the probability to find a photon in an
infinitesimal interval at frequency $\omega$, of the signal or idler
photon is given by
\begin{equation}
  \begin{split}
  S(\omega_s) &\propto \Abs{\infint[d\omega_i] f(\omega_s,
    \omega_i)}^2 \\
  S(\omega_i) &\propto \Abs{\infint[d\omega_s] f(\omega_s,
    \omega_i)}^2.
  \end{split}
\end{equation}
In the case of a highly coherent pump laser, 
$\alpha(\omega)$ can be approximated by a Dirac delta function,
$\delta(\omega - \omega_p)$, and the spectra of the signal and idler
photons is given by the phase matching, only, i.e.
\begin{equation}
  \label{eq:phasematching-sinc}
  S(\omega_j) \propto \text{sinc}^2(\Delta k L/2).
\end{equation}
The phase mismatch is given by
\begin{equation}
\label{eq:phase-mismatch}
\Delta k = 2\pi \left(\frac{n_p(\lambda_p)}{\lambda_p} -
  \frac{n_s(\lambda_s)}{\lambda_s} - \frac{n_i(\lambda_i)}{\lambda_i}
  - \frac{1}{\Lambda}\right),
\end{equation}
with $n_x$ and $\lambda_x\ (x = p, s, i)$ the refractive index and
wavelength of pump, signal and idler photons,
respectively. $\Lambda$ is the period of poling. Here, as a first
approximation, we have neglected the effect of the waveguide. A more
accurate expression would use the propagation constants of the pump,
signal and idler modes for the given waveguide refractive index profile.

We want to estimate the FWHM bandwidth of the photons generated by
SPDC\@. To this end, we first remember that $\lambda_i =
\left(\lambda_p^{-1} - \lambda_s^{-1}\right)^{-1}$ due to energy
conservation, such that the phase mismatch becomes a function of the
signal wavelength only. For phase-matching $\Delta k = 0$, and the
bandwidth is determined by the dispersion, which to first order is
given by
\begin{equation}
  \label{eq:deltak-approx}
  \begin{split}
    \Delta k(\lambda) &\simeq \frac{d \Delta k(\lambda)}{d \nu} \Delta\nu 
    = \frac{d \Delta k(\lambda)}{d \lambda}\frac{d \lambda}{d \nu}
    \Delta\nu \\
    &= \frac{d \Delta k(\lambda)}{d \lambda}\frac{- \lambda^2}{c}
    \Delta\nu \\
    &\equiv \Delta k' \Delta \nu.
  \end{split}
\end{equation}
Note that the contributions of the pump wavelength and the periodic
poling to $\Delta k(\lambda)$ are constant, so they will not affect
$\Delta k'$. Using Eq.~\eqref{eq:deltak-approx}, the argument of the
$\sinc^2$ function in Eq.~\eqref{eq:phasematching-sinc} becomes $x =
\Delta k' \Delta \nu L /2$. Knowing that the sinc squared reaches half
its maximum value at $x_{1/2} = 1.39156$, the FWHM bandwidh is given
by
\begin{equation}
\label{eq:fwhm-bandwidth}
\Delta \nu_\text{FWHM} = \frac{4 x_{1/2}}{\Abs{\Delta k'} L}
\end{equation}
Using the Sellmeier equations for KTP~\cite{Kato2002} and
LiNbO$_3$~\cite{Jundt1997}, we can calculate $\Delta k'$ and the
resulting values for $\Delta \nu_\text{FWHM}$. These are given in
Table~\ref{tab:fwhm-bandwidth}.

\begin{table}
  \caption{\label{tab:fwhm-bandwidth}Values for the estimation of the
    FWHM bandwidth of the two waveguides. For the PPLN waveguide we
    assume a temperature of \SI{180}{\celsius}.}
  \begin{ruledtabular}
    \begin{tabular}{lccc} 
      Waveguide & $\Delta k'$ (\si{(mm.GHz)^{-1}}) & $L$ (\si{mm}) 
      & $\Delta \nu_\text{FWHM}$ (\si{GHz}) \\ [0.2em]
      \hline \\ [-.5em]
      PPKTP & \num{-7.93e-4} & \num{13} & \num{539} \\
      PPLN & \num{-1.14e-3} & \num{50} & \num{97} \\[0.2em]
    \end{tabular}
  \end{ruledtabular} 
\end{table}

\section{Analytical model for SPDC with spectral filtering}
\label{app:pairsource-theory}
We shall here give a brief derivation of the expressions for the
signal and idler rates, the coincidence rate and the second-order
correlation function of the waveguides, including the application of
spectral filtering. As a starting point we will take the treatment
presented by Razavi et al.~\cite{Razavi2009} (see
also~\cite{Wong2006}), assuming collinear SPDC with plane-wave
fields. Furthermore, the depletion of the pump and group-velocity
dispersion have been neglected.

We start by giving expressions for the first-order correlation
functions, from which one can calculate the event rates. With the help
of the quantum form of the Gaussian moment-factoring theorem, all
higher-order correlation functions can be derived \cite{Razavi2009}.

\subsection{First-order correlation functions}
\label{app:first-order-correlations}
Defining scalar photon-units positive-frequency field operators,
\begin{equation}
  \label{eq:field-operators}
  E_j(t) = \frac{1}{2\pi}\infint[d\omega] a(\omega) e^{-i\omega t},
  \qquad j=s,i,
\end{equation}
where $a(\omega)$ is the photon annihilation operator in the frequency domain,
Razavi et al\@. use a Bogoliubov transformation to derive the
following set of first-order correlation functions for the SPDC output state,
\begin{equation}
  \label{eq:correlations-razavi}
  \begin{split}
  \expect{E_j^\dag(t+\tau)E_j(t)} &=
  e^{i\omega_j \tau} \times  C_\text{auto}(\tau) \\
  \expect{E_j(t+\tau)E_k(t)} &=
  \begin{aligned}[t]
  (1-\delta_{jk}) e^{-i(\omega_p t + \omega_j \tau)} \\
  \times C_\text{cross}(\tau),
  \end{aligned}
\end{split}
\end{equation}
where $\delta_{jk}$ is the Kronecker delta function and $j,k \in
\{s,i\}$. In the low-gain regime of SPDC, the envelope functions
$C_\text{auto}(\tau)$ and $C_\text{cross}(\tau)$ are 
given by
\begin{equation}
  \label{eq:correlation-envelopes}
  \begin{split}
    C_\text{auto}(\tau) &=
    \begin{cases} 
      R\left(1-\abs{\tau}
        B\right) & \text{for } \abs{\tau} B \leq 1 \\
      0 & \text{otherwise}
    \end{cases}, \\
    C_\text{cross}(\tau) &=
    \begin{cases}
      \sqrt{R B} & \text{for } \abs{\tau} B \leq \frac{1}{2} \\
      0 & \text{otherwise}
    \end{cases}.
  \end{split}
\end{equation}
Here, $R$ is the rate of photon pair creation and proportional
to the pump power, and $B=2\pi/(\Delta k' L)$ is proportional to the
bandwidth. The ratio $R/B$ is often termed the spectral brightness of
the photon pair source.

When adding spectral filtering, the envelope functions get
convoluted with the impulse response functions $F_j(t)$ of the
filters~\cite{Mitchell2009}. For the autocorrelation, 
\begin{equation}
  \label{eq:correlations-filtered}
  \begin{split}  
    C_\text{auto}^{(j)}(\tau) &=
    \begin{aligned}[t]
      \infint[dt'] \!\!\!\! \infint[dt'']
      &F_j^*(t+\tau-t') F_j(t-t'') \\ & \times C_\text{auto}(t' - t'')
    \end{aligned} \\
    &\approx \alpha \infint[dt'] F_j^*(t+\tau-t') F_j(t-t'),
  \end{split}
\end{equation}
where we have taken $C_\text{auto}(t' - t'') \approx \alpha\,
\delta(t'-t'')$, which is valid if the bandwidth of the filter is much
smaller than $B$. The constant $\alpha$ is
\begin{equation}
  \label{eq:alpha}
  \alpha = \infint\!\!\!dt' C_\text{auto}(t'-t'') = \frac{R}{B}.
\end{equation}
We further consider a Lorentzian filter with FWHM $\Gamma_j$ whose
transfer and impulse response functions are given by
\begin{align}
  \label{eq:lorentzian-transfer}
  H_j(\omega) & = \frac{\Gamma_j}{\Gamma_j - 2i\omega} \\
  \label{eq:lorentzian-response}
  \begin{split}
    F_j(\tau) &= \frac{1}{2\pi}\infint[d\omega]
    H_j(\omega)\, e^{i\omega\tau}\\
    & = \frac{\Gamma_j}{2} \Theta(\tau) e^{-\Gamma_j \tau/2},
  \end{split}
\end{align}
where $\Theta(\tau)$ is the Heaviside step function. We then arrive at
the final expression for the auto-correlation envelope,
\begin{equation}
  \label{eq:auto-correlation-final}
  C_\text{auto}^{(j)}(\tau) = 
  \frac{1}{4} \frac{R}{B} \Gamma_j e^{-\Gamma_j \abs{\tau}/2}.
\end{equation}

Performing a similar calculation for the cross-correlation
envelope, we get
\begin{equation}
  \label{eq:cross-correlation-final}
  C_\text{cross}^{(jk)}(\tau) = \frac{1}{2} \sqrt{\frac{R}{B}} \frac{\Gamma_j
    \Gamma_k}{\Gamma_j + \Gamma_k} \times
  \begin{cases}
    e^{\Gamma_k \tau/2} & \text{for } \tau < 0 \\
    e^{-\Gamma_j \tau/2} & \text{for } \tau \geq 0
  \end{cases}.
\end{equation}

Finally, let us introduce, for convenience, the signal and idler flux,
\begin{equation}
  \label{eq:singles-flux}
  W_j \equiv C_\text{auto}^{(j)}(0) = \frac{1}{4}\frac{R}{B}\Gamma_j,
\end{equation}
and the pair flux,
\begin{equation}
  \label{eq:pair-flux}
  \begin{split}
    W_2 &\equiv \infint[d\tau]
    \Abs{C_\text{cross}^{(jk)}(\tau)}^2 \\
    & = \frac{1}{4} \frac{R}{B} \frac{\Gamma_j \Gamma_k}{\Gamma_j +
      \Gamma_k} \\
    & = W_j \times \frac{\Gamma_k}{\Gamma_j + \Gamma_k}.
  \end{split}
\end{equation}
The last line of Eq.~\eqref{eq:pair-flux} says that the pair flux is
equal to the flux if signal or idler rescaled by the probability that
a photon that has already been projected onto the spectrum of one of
the filters also passes the second filter. We note that this
expression is valid only for perfectly correlated photon pairs and
does not contain contributions from multi-pair emission. These will be
included in the next section, where we consider second-order
correlation functions.

\subsection{Second-order correlation functions}
\label{app:2nd-order-correlations}
The normalized second-order cross-correlation function is defined as
\begin{equation}
  \label{eq:g2si-definition}
  \begin{split}
  g_{si}^{(2)}(\tau) &\equiv \frac{\expect{E_s^\dag(t) E_i^\dag(t+\tau)
      E_i(t+\tau) E_s(t)}}{\expect{E_s^\dag(t)
      E_s(t)}\expect{E_i^\dag(t+\tau) E_i(t+\tau)}} \\
  &= \frac{G_{si}^{(2)}(\tau)}{W_s\, W_i},
\end{split}
\end{equation}
where the numerator is the non-normalized
second-order cross-correlation function. Applying the Gaussian
moment-factoring theorem, it can be shown that
\begin{equation}
  \label{eq:G2si}
  G_{si}^{(2)}(\tau) = W_s\, W_i + \Abs{C_\text{cross}^{(si)}(\tau)}^2,
\end{equation}
where the first term is proportional to the coincidence rate that is
expected for completely uncorrelated photons, often called accidental
coincidences.  Using Eqs.~\eqref{eq:cross-correlation-final} and
\eqref{eq:singles-flux}, we find
\begin{equation}
  \label{eq:g2si}
  \begin{split}
  g_{si}^{(2)}(\tau) &= 1 +
  \frac{\Abs{C_\text{cross}^{(si)}(\tau)}^2}{W_s W_i} \\
  &= 1 + 4\frac{B}{R} \frac{\Gamma_s \Gamma_i}{(\Gamma_s +
    \Gamma_i)^2} \times f_{si}(\tau) \\
  f_{jk}(\tau) &=
  \begin{cases}
    e^{\Gamma_j \tau} & \text{for } \tau < 0 \\
    e^{-\Gamma_k \tau} & \text{for } \tau \geq 0
  \end{cases}.
\end{split}
\end{equation}

The derivation of the second-order auto-correlation functions for the
signal and idler photons proceeds along the same lines as that of the
cross-correlation. The auto-correlation function is defined as
\begin{equation}
  \label{eq:g2jj-definition}
  g_{jj}^{(2)}(\tau) \equiv \frac{\expect{E_j^\dag(t) E_j^\dag(t+\tau)
      E_j(t+\tau) E_j(t)}}{\expect{E_j^\dag(t)
      E_j(t)}\expect{E_j^\dag(t+\tau) E_j(t+\tau)}}.
\end{equation}
Applying the same steps as before, this can be shown
to be equal to
\begin{equation}
  \label{eq:g2-final}
  g_{jj}^{(2)}(\tau) = 1 +
  \frac{\abs{C_\text{auto}^{(j)}(\tau)}^2}{W_j^2} = 1 + f_{jj}(\tau),
\end{equation}
where we have reused the definition of $f_{jk}(\tau)$ from
Eq.~\eqref{eq:g2si}. 

\subsection{Inclusion of experimental imperfections}
Before the expressions derived in the
appendices~\ref{app:first-order-correlations}
and~\ref{app:2nd-order-correlations} can be compared to the
experimental data, they need to be slightly modified to take into
account experimental imperfections in the shape of finite
efficiencies, dark counts and electronic jitter.

Let us start by considering the jitter of our detection system, which
is well modeled by a normal distribution
\begin{equation}
  \label{eq:jitter}
  j(t) = \frac{1}{\sqrt{2\pi \sigma^2}} e^{-t^2/(2 \sigma^2)}.
\end{equation}
The effect on the measured cross- and auto-correlation functions
can be calculated as the convolution of $f_{jk}(\tau)$ from
Eq.~\eqref{eq:g2si} with $j(t)$, and one obtains
\begin{equation}
  \label{eq:jitter-lineshape}
  \begin{split}
    \tilde{f}_{jk}(\tau) = \frac{1}{2}
      & \left[ e^{\Gamma_j(\Gamma_j \sigma^2/2 +t)}
      \erfc\left(\frac{\Gamma_j\,\sigma^2 + t}{\sqrt{2}\,\sigma}
      \right) \right. \\
      & \left. + e^{\Gamma_k(\Gamma_k \sigma^2/2 - t)}
      \erfc\left(\frac{\Gamma_k\,\sigma^2 - t}{\sqrt{2}\,\sigma}
      \right) \right]
  \end{split}
\end{equation}

The spectral filters do not have unit peak
transmission. Additionally, the detectors have a finite efficiency and
there is loss on the surfaces of optical elements and when coupling
into single-mode fiber. By gathering all the losses into a single
coefficient, they can be taken into account by adding a prefactor of
$\sqrt{\eta_j}$ to the transfer
function~\eqref{eq:lorentzian-transfer}. This leads to a reduction of
the signal and idler flux~\eqref{eq:singles-flux} by a factor of
$\eta_j$, and the pair flux~\eqref{eq:pair-flux} is correspondingly
reduced by a factor $\eta_j \eta_k$.

Besides the finite efficiency of the filtering, the etalon or cavity
may not be well-approximated by a single Lorentzian filter. This is
the case if more than one longitudinal mode is excited. Spurious modes
contribute the photon flux and increase it by a factor $1/p_0$ where
$p_0$ is fraction of the photons that end up in the desired
mode. However, spurious modes cannot contribute to the pair flux,
since the free spectral ranges of etalon and cavity are incommensurate.
As explained in the main text, the signal filtering suffers from such
spurious modes, and a correction has been added to the signal flux.

Detector dark counts add an offset to the detected photon
flux and will also contribute to the accidental coincidences. This
effect can be added to the formalism by introducing a constant term
$D_j$ to Eq.~\eqref{eq:singles-flux} and using
Eqs.~\eqref{eq:g2si-definition} and~\eqref{eq:G2si} for comparison
with the measurements, instead of the simplified
expression~\eqref{eq:g2si}. Please note that the pair flux $W_2$ by
definition does not contain contributions from accidental coincidences.
In summary, the experimental data presented in
Fig.~\ref{fig:characterization} has been fitted to the expressions
\begin{equation}
  \label{eq:app:characterization-fit}
  \begin{split}
    W_s &= \frac{1}{4}\frac{\eta_s}{p_0} \frac{R}{B} \Gamma_s + D_s \\
    W_i &= \frac{1}{4}\eta_i \frac{R}{B} \Gamma_i + D_i \\
    W_2 & = \frac{1}{4} \eta_s \eta_i \frac{R}{B} \frac{\Gamma_s
      \Gamma_i}{\Gamma_s + \Gamma_i} \\
    g_{si}^{(2)}(0) &= 1 + \frac{1}{4} \frac{\tilde{f}_{si}(0)\, \eta_s \eta_i}{W_s
      W_i} \frac{R}{B} \left( \frac{\Gamma_s \Gamma_i}{\Gamma_s +
        \Gamma_i} \right)^2               
  \end{split}
\end{equation}
with the free parameters $\eta_s, \eta_i, R/B$.

\section{Details for the violation of the CHSH inequality}
The violation of the CHSH inequality requires the joint measurement of the
signal and idler photons in four combinations of bases. In our case,
we chose the idler bases $X_1$ and $X_2$ to correspond to the Pauli
matrices $\sigma_x$ and $\sigma_y$, respectively. If the source would
produce the Bell state $\ket{\Phi^+}$,
i.e. Eq.~(\ref{eq:max-entangled-pair}) with $\phi=0$, an optimal
choice for the signal photon could be $Y_{1,2}=(\sigma_x \pm
\sigma_y)/\sqrt{2}$. For non-zero $\phi$, this can be generalized to
$Y_{1,2} = \cos\theta_\pm\sigma_x + \sin \theta_\pm \sigma_y$ with
$\theta_\pm = \phi \pm \sfrac{\pi}{4}$. In the experiment, we first
determined $\phi$ by a separate measurement and then proceeded to the
violation of the CHSH inequality, which consists of measuring the four
correlators
\begin{equation}
  \label{eq:chsh-correlator}
  E(X_i, Y_i) = \frac{N_{11} + N_{22} - N_{12} - N_{21}}{N_{11} + N_{22} + N_{12} + N_{21}},
\end{equation}
where, e.g., $N_{11}$ is the number of coincidences between detectors Si$_1$
and ID220$_1$. The CHSH parameter is then given by
\begin{equation}
  \label{eq:chsh-parameter}
  S = \abs{E(X_1, Y_1) + E(X_1, Y_2) + E(X_2, Y_1) - E(X_2, Y_2)}.
\end{equation}
We obtained the following values for the correlators,
\begin{align*}
  \label{eq:chsh-correlators-measured}
  E(X_1, Y_1) &= 0.638(5)\\
  E(X_1, Y_2) &= 0.702(5)\\
  E(X_2, Y_1) &= 0.700(5)\\
  E(X_2, Y_2) &= -0.669(5)\\
\end{align*}
which gives $S = 2.708(9)$.

\bibliography{photonsource}

\end{document}